\documentclass[10pt,letterpaper]{article}
\usepackage[top=0.85in,left=2.75in,footskip=0.75in]{geometry}

\usepackage{changepage}

\usepackage[utf8]{inputenc}

\usepackage{textcomp,marvosym}

\usepackage{fixltx2e}

\usepackage{amsmath,amssymb}

\usepackage{cite}

\usepackage{nameref,hyperref}

\usepackage[right]{lineno}

\usepackage{microtype}
\DisableLigatures[f]{encoding = *, family = * }

\usepackage{rotating}

\raggedright
\usepackage[aboveskip=1pt,labelfont=bf,labelsep=period,justification=raggedright,singlelinecheck=off]{caption}

\makeatletter
\renewcommand{\@biblabel}[1]{\quad#1.}
\makeatother

\date{}

\usepackage{lastpage,fancyhdr,graphicx}
\usepackage{epstopdf}
\fancyheadoffset[L]{2.25in}
\fancyfootoffset[L]{2.25in}

\begin{document}
\vspace*{0.35in}

\begin{flushleft}
{\Large
\textbf\newline{Different Evolutionary Paths to Complexity for Small and Large Populations of Digital Organisms}
}
\newline
\\
Thomas LaBar\textsuperscript{1,2,3,*},
Christoph Adami\textsuperscript{1,2,3,4}
\\
\bigskip
\bf{1} Department of Microbiology and Molecular Genetics, Michigan State University, East Lansing, MI, United States of America
\\
\bf{2} Ecology, Evolutionary Biology, and Behavior Program, East Lansing, MI, United States of America
\\
\bf{3} BEACON Center for the Study of Evolution in Action, Michigan State University, East Lansing, MI, United States of America
\\
\bf{4} Department of Physics and Astronomy, Michigan State University, East Lansing, MI, United States of America
\\
\bigskip

* labartho@msu.edu

\end{flushleft}

\section*{Abstract}
A major aim of evolutionary biology is to explain the respective roles of adaptive versus non-adaptive changes in the evolution of complexity. While selection is certainly responsible for the spread and maintenance of complex phenotypes, this does not automatically imply that strong selection enhances the chance for the emergence of novel traits, that is, the origination of complexity. Population size is one parameter that alters the relative importance of adaptive and non-adaptive processes: as population size decreases, selection weakens and genetic drift grows in importance. Because of this relationship, many theories invoke a role for population size in the evolution of complexity. Such theories are difficult to test empirically because of the time required for the evolution of complexity in biological populations. Here, we used digital experimental evolution to test whether large or small asexual populations tend to evolve greater complexity. We find that both small and large---but not intermediate-sized---\-populations are favored to evolve larger genomes, which provides the opportunity for subsequent increases in phenotypic complexity. However, small and large populations followed different evolutionary paths towards these novel traits. Small populations evolved larger genomes by fixing slightly deleterious insertions, while large populations fixed rare beneficial insertions that increased genome size. These results demonstrate that genetic drift can lead to the evolution of complexity in small populations and that purifying selection is not powerful enough to prevent the evolution of complexity in large populations.

\section*{Author Summary}
Since the early days of theoretical population genetics. scientists have debated the role of population size in shaping evolutionary dynamics. Do large populations possess an evolutionary advantage towards complexity due to the strength of natural selection in these populations? Or do small populations have the advantage, as genetic drift allows for the exploration of fitness landscapes inaccessible to large populations? There are many theories that predict whether large or small populations--those with strong selection or those with strong drift--should evolve the greatest complexity. Here, we use digital experimental evolution to examine the interplay between population size and the evolution of complexity. We found that genetic drift could lead to increased genome size and phenotypic complexity in very small populations. However, large populations also evolved similar large genomes and complexity. Small populations evolved larger genomes through the fixation of slightly deleterious insertions, while large populations utilized rare beneficial insertions. Our results suggest that both strong drift and strong selection can allow populations to evolve similar complexity, but through different evolutionary trajectories.
\nolinenumbers

\section*{Introduction}
The relative importance of adaptive (i.e., selection) versus non-adaptive (i.e., drift) mechanisms in shaping the evolution of complexity is still a matter of contention among evolutionary biologists~\cite{bonner1988evolution,adami2000evolution,koonin2004non,lynch2007frailty,tenaillon2007quantifying,mcshea2010biology}. In molecular evolution, the role of non-adaptive evolutionary processes such as genetic drift and genetic draft are well-established~\cite{kimura1984neutral,ohta1992nearly,gillespie2000genetic}. Theoretical population-genetic principles argue that neutral evolution, not natural selection, drove the evolution of large, primarily non-functional, genomes~\cite{lynch2011repatterning,eddy2012c,palazzo2014case}. Meanwhile, there exists abundant experimental evidence that natural selection is the main cause of evolutionary change~\cite{travisano1995experimental,wagenaar2004influence,lachapelle2015repeatability}, including the spread of novel adaptive phenotypes~\cite{blount2008historical,meyer2012repeatability}, in experimental populations. However, it is still possible that non-adaptive processes play a significant role in the evolution of complexity. For instance, genetic drift, or relaxed selection, may allow for the accumulation of mutations that can later lead to the evolution of novel complexity~\cite{lynch2007frailty,wagner2011origins}. Much of the work demonstrating the role of selection in driving the evolution of novel complex traits is based on experiments with large populations and strong selection~\cite{kawecki2012experimental}. In much smaller populations (i.e., those with fewer than $10^4$ individuals), selection is weaker, and genetic drift begins to alter evolutionary dynamics~\cite{masel2011genetic,lachapelle2015repeatability}. Therefore, to explain the role of adaptive vs. non-adaptive process in the evolution of complexity, one must explore the role of population size in the evolution of complexity.

Both theoretical modeling and experiments suggest many possibilities for the relationship between population size and the evolution of complexity. There are two classes of evolutionary trajectories that would favor large populations in the evolution of complexity. First, populations could perform an adaptive walk (the fixation of a sequence of beneficial mutations) towards the evolution of a novel complex trait~\cite{schoustra2009properties}. If this was the case, then larger populations would follow this trajectory faster than small populations due to their larger mutation supply. Experiments with microorganisms support the possible existence of adaptive trajectories towards complexity, as there is strong evidence that the mutations leading up to a phenotypic innovation in both \textit{Escherichia coli}~\cite{quandt2015fine} and phage $\lambda$~\cite{burmeister2015selection} were under positive selection. However, it is unclear whether adaptive mutations generally precede the evolution of complex traits or whether these large microbial populations can only take adaptive walks due to the intensity of selection in large populations. The second type of trajectory that favors large populations is the neutral walk (the fixation of a sequence of neutral mutations). While any individual neutral mutation has a low probability of fixation, a large population would be able to accumulate many neutral mutations at any given time allowing for the exploration of its fitness landscape. Work by Wagner and colleagues suggests that many phenotypic traits are connected to each other by sequences of phenotypically neutral mutations~\cite{wagner2008neutralism,wagner2011origins}. 

If the evolution of complexity requires the fixation of deleterious mutations (for example, via valley crossing), then the elimination of deleterious mutations by purifying selection may limit the evolutionary advantage large populations may have. Wright was the first to propose an evolutionary advantage of small populations due to valley-crossing~\cite{wright1932roles}. More recently, scientists have explored under which conditions small populations have an evolutionary advantage over large populations~\cite{rozen2008heterogeneous,jain2011evolutionary}. A prominent theory that predicts that small (but not large) populations should evolve the greatest genomic complexity (and subsequently organismal complexity) is the Mutational Burden (or Mutational Hazard) hypothesis, proposed by Lynch and colleagues~\cite{lynch2003origins,lynch2007origins,lynch2007frailty}. This hypothesis argues that genome size should be inversely correlated with the product of the effective population size and the mutation rate~\cite{lynch2003origins,koonin2004non}. Strong purifying selection against excessive genome size streamlines the genomes in large populations~\cite{lynch2006streamlining,batut2014reductive,zwart2014experimental}. Meanwhile, weakened purifying selection and increased genetic drift in small populations results in the accumulation of slightly deleterious excess genome content~\cite{koonin2004non,lynch2007origins}. At a later time, this slightly deleterious genome content may be mutated into novel beneficial traits~\cite{lynch2007frailty,koonin2012logic}. However, recent work on valley crossing in asexual populations (and sexual populations with a low recombination rate) showed that both small and large populations valley-cross more than intermediate-sized populations~\cite{weissman2009rate,weissman2010rate,ochs2015competition}. Therefore, it is not clear whether large or small populations are expected to evolve the greatest complexity when deleterious mutations are required.

The long timescales required to observe the emergence of novelty and evolution of complexity make biological experiments to distinguish between these theories difficult to perform. To overcome this difficulty, we used digital experimental evolution~\cite{adami2006digital} to test the role of population size on the evolution of genome size and phenotypic complexity in asexual organisms. Digital evolution has a long history of addressing macroevolutionary questions (such as the evolution of novel traits) experimentally~\cite{yedid2002macroevolution,bell2016experimental}. Digital evolution makes it possible to manipulate an evolving population in ways populations of biochemical organisms can not, in order to test which factors result in certain evolutionary outcomes~\cite{batut2013silico}. In this regard, digital experimental evolution has the same goals as microbial experimental evolution: to use a well-controlled model system that is as simple as possible, to study ``evolution in action"~\cite{elena2003evolution}. And while digital evolution studies cannot test hypotheses dependent on particular biochemical processes involved in cellular life, digital populations do undergo selection, drift, and mutation, allowing for their use in testing hypotheses derived from theoretical population genetics. Thus, digital experimental evolution represents a well-suited model system to test the population genetics-based theories concerning the role of population size in the evolution of complexity.

Here, we evolved populations ranging in size from $10$ to $10^4$ individuals, starting with a minimal genome ancestor. We found that small populations do evolve greater genome sizes and hence phenotypic complexity than intermediate-sized populations. These small populations evolve larger genomes primarily through increased fixation of slightly deleterious insertions. However, the small population sizes that enhance the evolution of phenotypic complexity also enhance the likelihood of population extinction. We also found that the largest populations evolved similar complexity to the smallest populations. Large populations evolved longer genomes and greater phenotypic complexity through the fixation of rare beneficial insertions instead. Large populations were able to discover these rare beneficial mutations due to an increased mutation supply.  Finally, we found that a strong deletion bias can prevent the evolution of greater complexity in small, but not in large, populations.    

\section*{Results}
To explore the effect of population size on the evolution of genome size and phenotypic complexity, we use the Avida digital evolution system~\cite{ofria2009avida}. Avida is a platform that allows researchers to perform evolution experiments inside of a computer, as the genetic code that evolves are actual computer programs of variable length. It has been used extensively in research in evolutionary biology~\cite{Adami1998,WilkeAdami2002,adami2006digital}, and is described in detail in Methods.

We evolved one hundred replicate populations across a range of population sizes ($10-10^4$ individuals) for $2.5\times10^5$ generations. Many of the smallest populations (those with ten individuals) did not survive the entire experiment. Therefore, we evolved one hundred additional small populations ranging from twenty individuals to ninety individuals in order to examine how the probability of extinction was related to the evolution of complexity. All populations with at least thirty individuals survived for the entire experiment. Forty-seven of the populations with ten individuals went extinct, while only one of one hundred populations underwent extinction in the populations with twenty individuals. Extinction was a consequence of populations evolving large genomes that accumulated deleterious mutations and led to the production of only non-viable offspring. These extinct populations were not included in the statistics described below.

\subsection*{Genome Size Evolution}
\begin{figure}[!h]
\includegraphics[width=4in]{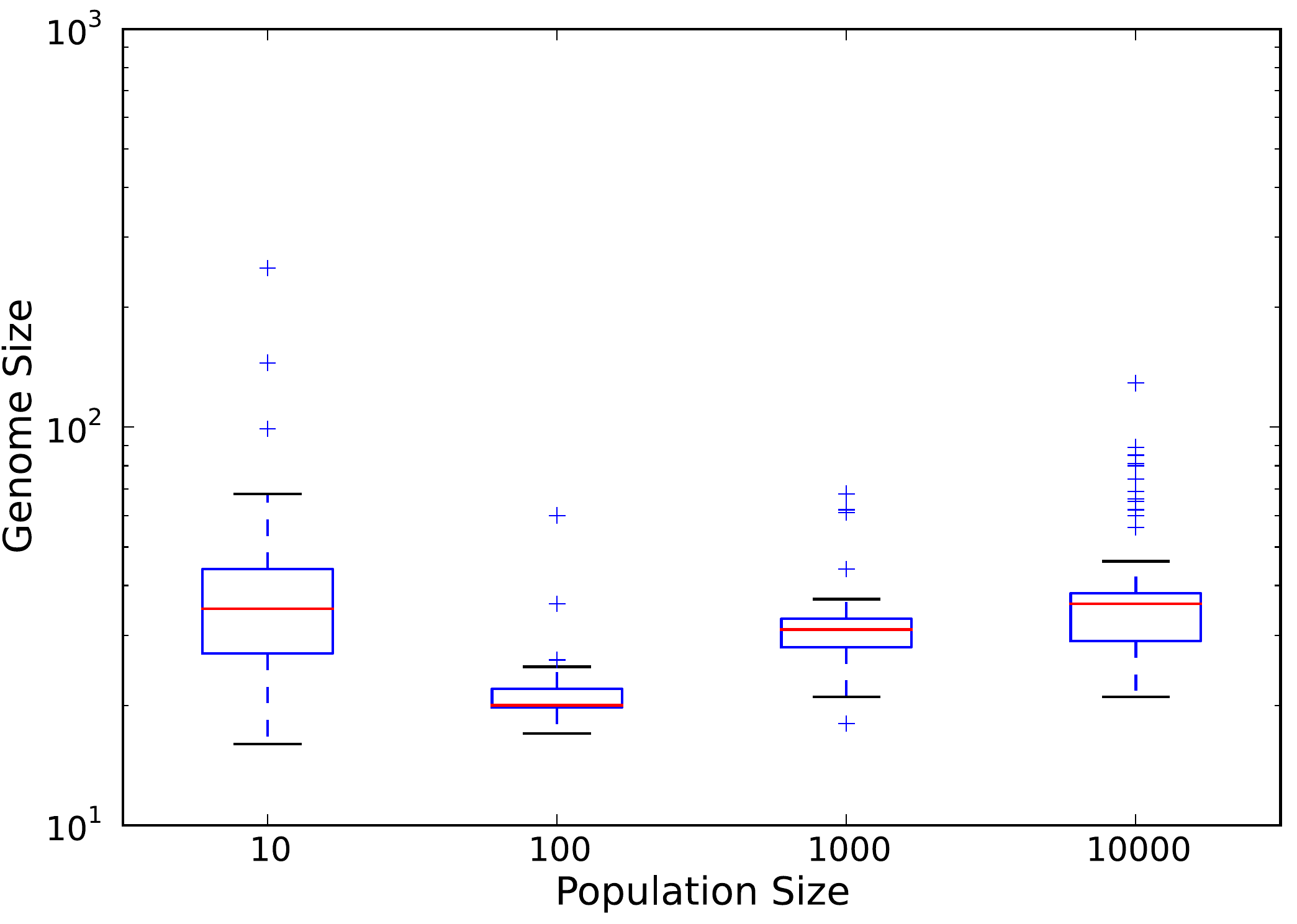} 
\caption{{\bf Final genome size as a function of population size.} Red lines are the median values for each population size. The upper and lower limits of each box denote the third and first quartile, respectively. Whiskers are 1.5 times the relevant quartile value. Plus signs denote those data points beyond the whiskers. Data represent only those populations that did not go extinct.}
\label{fig1}
\end{figure}

Of the surviving populations, we first examined how genome size changes from the ancestral value of fifteen instructions. The size of the genome from every population size increased, on average (see Fig.~\ref{fig1} and panel A in Fig. S1). However, both the smallest and the largest populations evolved the largest genomes. Populations with ten individuals evolved a median genome size of 35 instructions, while populations with ten thousand individuals evolved a median genome size of 36 instructions. The median final genome size decreased as population size increased for populations with between ten and fifty individuals. However, from populations with fifty individuals to populations with ten thousand individuals, the median final genome size increased as population size increased.

\begin{figure}[!h]
\includegraphics[width=4in]{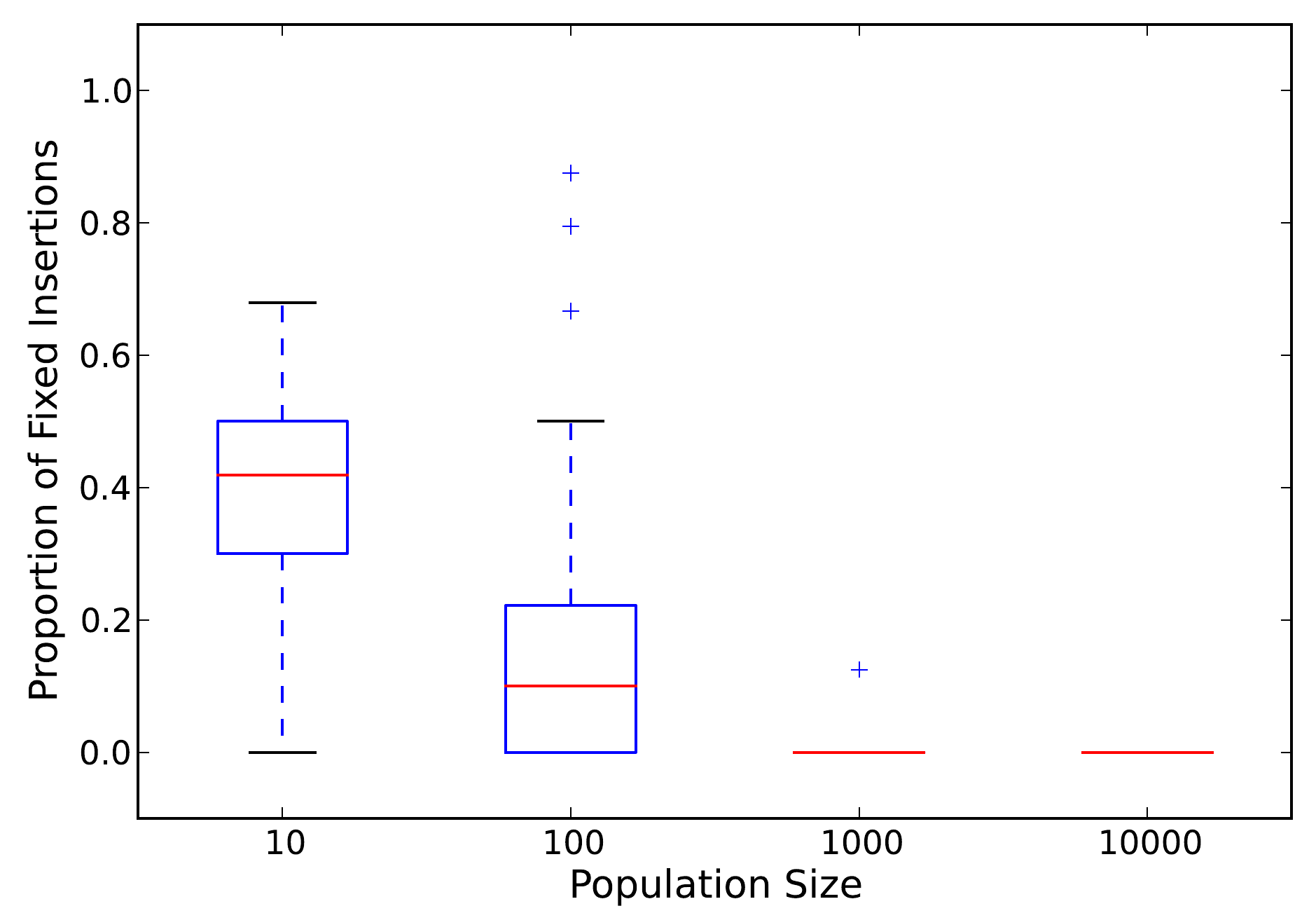} 
\caption{{\bf Proportion of slightly-deleterious insertions as a function of population size.}
Red lines are the median values for each population size. The upper and lower limits of each box denote the third and first quartile, respectively. Whiskers are 1.5 times the relevant quartile value. Plus signs denote those data points beyond the whiskers. Data represent only those populations that did not go extinct.}
\label{fig2}
\end{figure}

\begin{figure}[!h]
\includegraphics[width=4in]{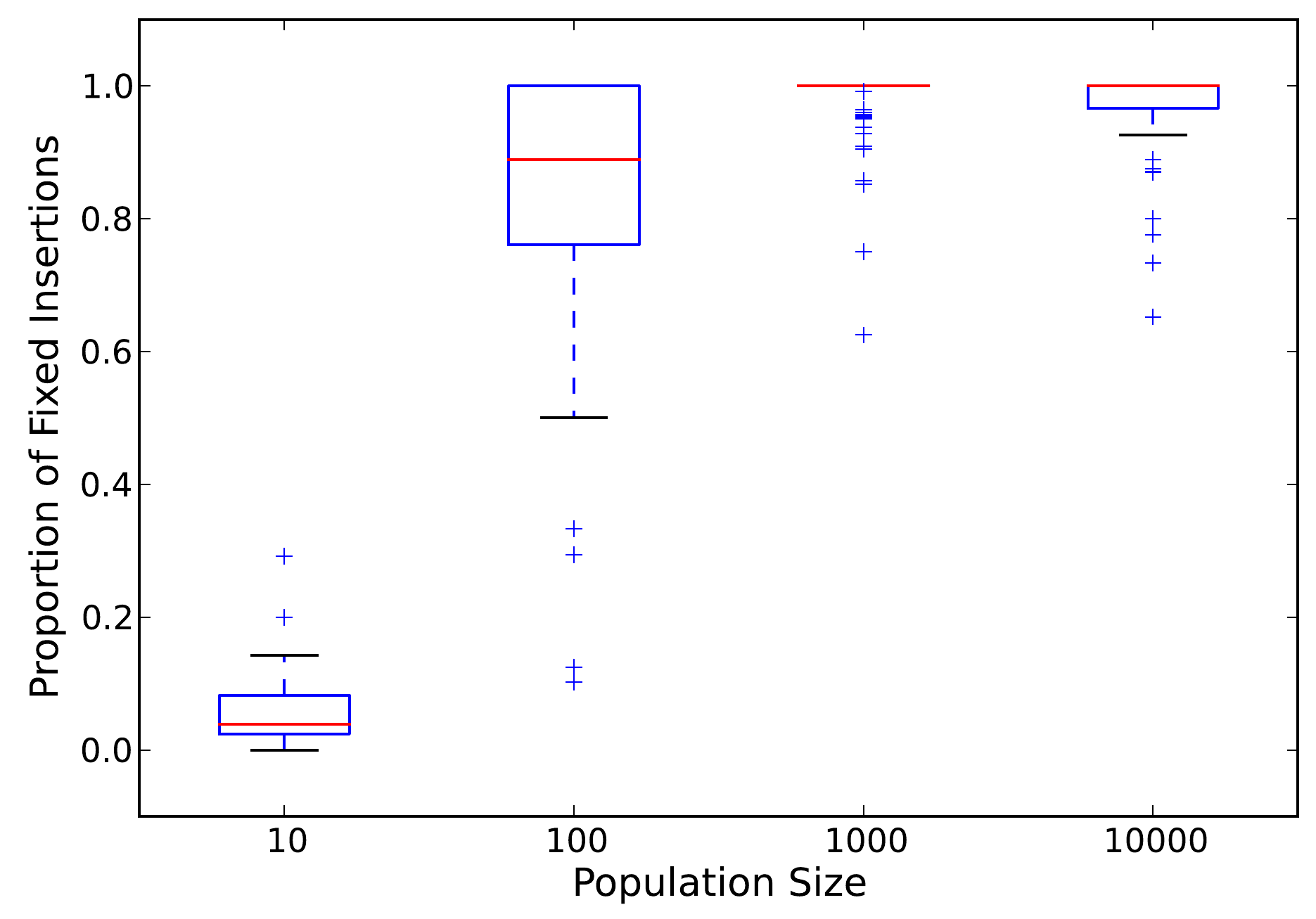} 
\caption{{\bf Proportion of insertions under positive selection as a function of population size}
Red lines are the median values for each population size. The upper and lower limits of each box denote the third and first quartile, respectively. Whiskers are 1.5 times the relevant quartile value. Plus signs denote those data points beyond the whiskers. Data represent only those populations that did not go extinct.}
\label{fig3}
\end{figure}

Next, we examined the dynamics of fixation of insertion mutations (insertions, for short) to explain why both the smallest and the largest populations evolved the largest genomes. For each experimental population, we counted every insertion that occurred on the fittest genotype's ancestral lineage that went back to the ancestral genotype (the ``line of descent", see Methods). The median number of insertions fixed follows the same trend as the evolution of genome size (Fig. S2). A large fraction of these fixed insertions are slightly deleterious in populations with fewer than one hundred individuals (see Fig.~\ref{fig2} and panel B in Fig. S1). However, no insertions are slightly deleterious, on average, in large populations with more than one hundred individuals. The opposite trend holds for beneficial insertions. The fraction of insertions that are under positive selection increases with increasing population size, with the largest populations usually fixing only beneficial insertions (Fig.~\ref{fig3} and panel C in Fig. S1). These data demonstrate that small populations evolve larger genomes  through the fixation of slightly deleterious insertions. However, large populations can evolve similarly large genomes through the fixation of rare beneficial insertions.

\subsection*{Evolution of Phenotypic Complexity}
\begin{figure}[!h]
\includegraphics[width=4in]{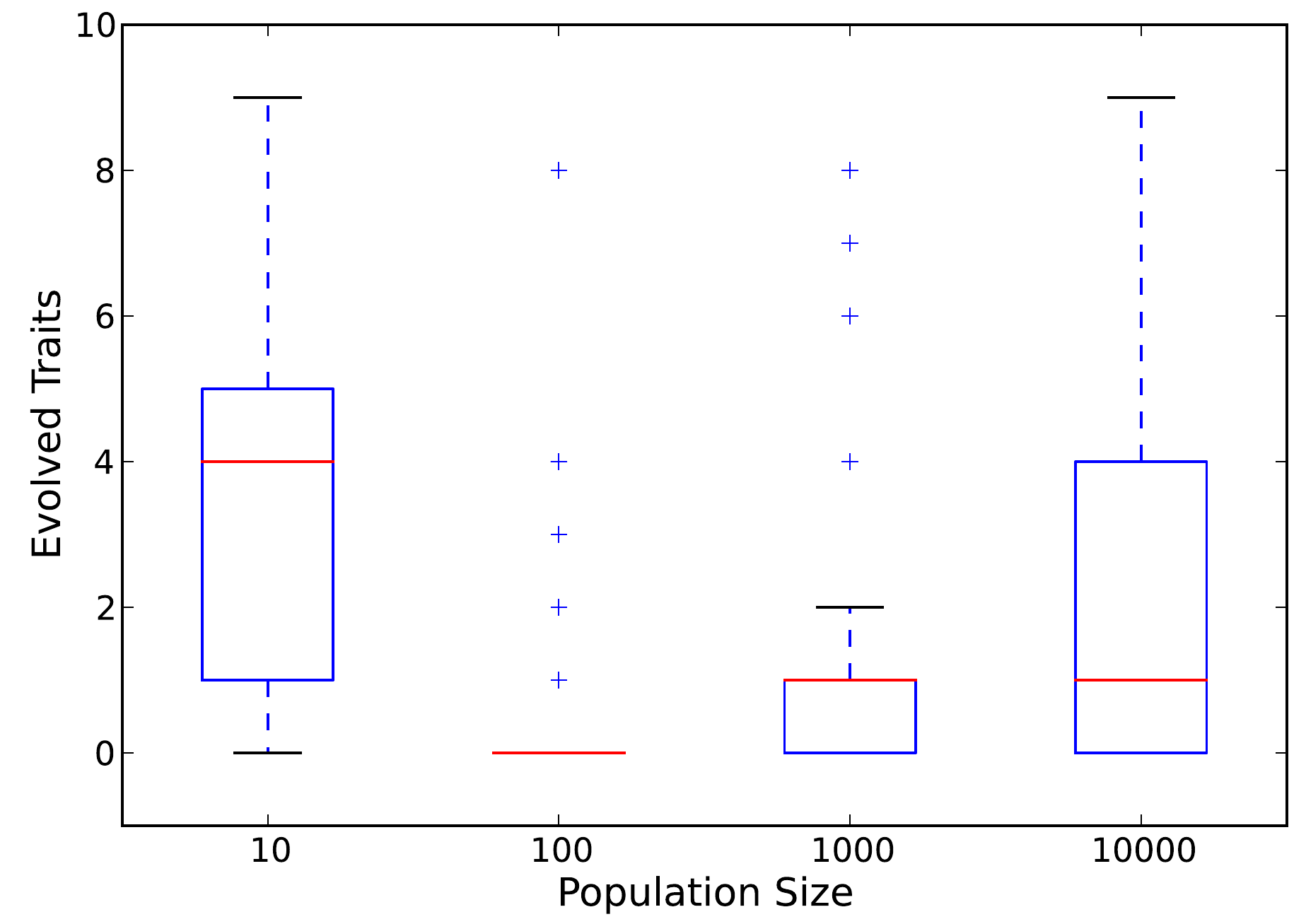}
\caption{{\bf Final number of evolved phenotypic traits as a function of population size.}
Red lines are the median values for each population size. The upper and lower limits of each box denote the third and first quartile, respectively. Whiskers are 1.5 times the relevant quartile value. Plus signs denote those data points beyond the whiskers. Data represent only those populations that did not go extinct.}
\label{fig4}
\end{figure}

Next, we focus on the role of population size in the evolution of phenotypic complexity (defined as the number of phenotypic traits). In Avida, a phenotypic trait is a program's ability to perform a certain mathematical operation on binary numbers (see Methods). The evolution of phenotypic complexity follows the same trend as the evolution of genome size (see Fig.~\ref{fig4} and panel D in Fig. S1). Populations with ten individuals evolved a median of four traits, while populations with one thousand and ten thousand individuals evolved a median of one trait. The rest of the population sizes evolved a median of zero traits.

\begin{figure}[!h]
\includegraphics[width=4in]{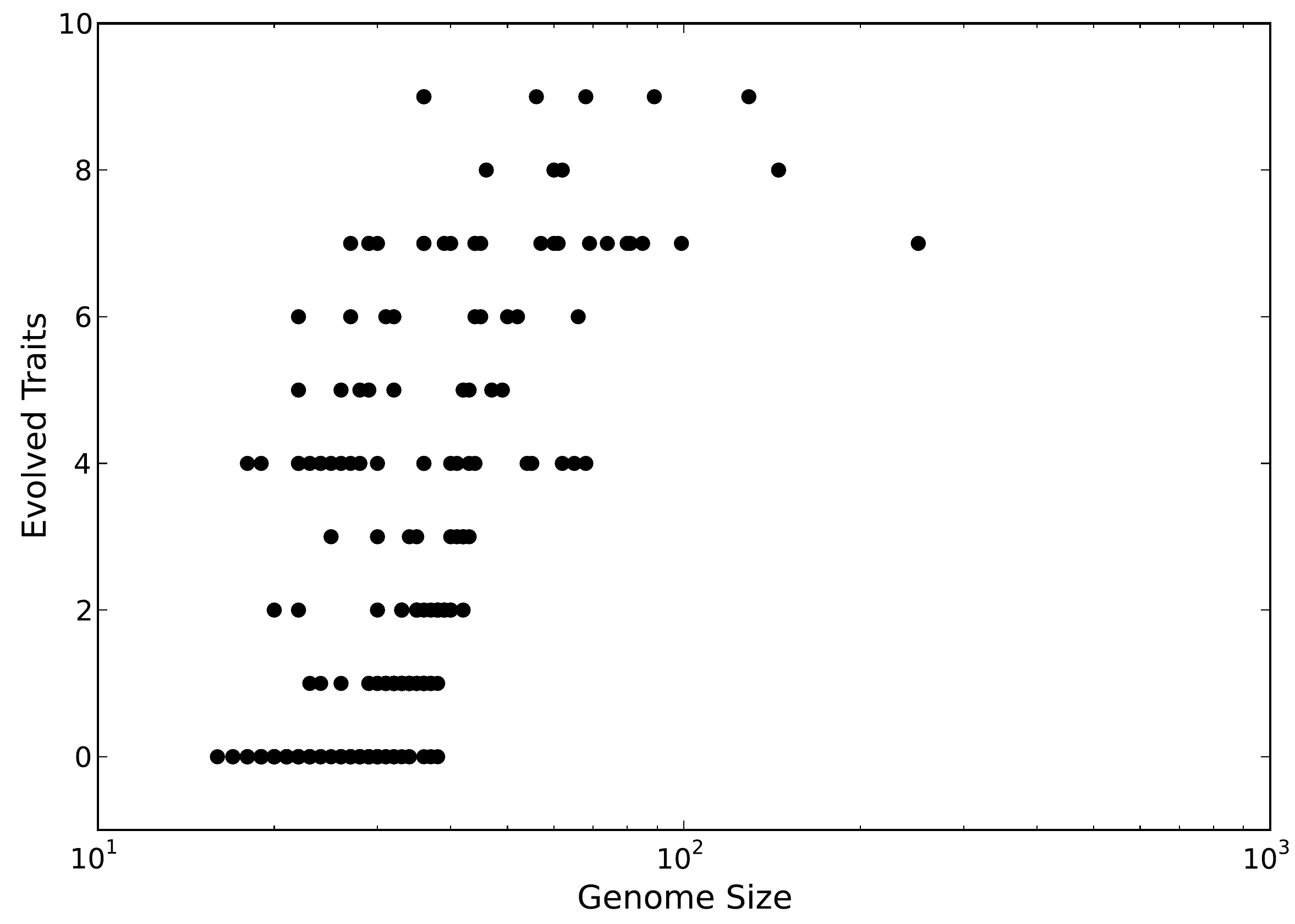}
\caption{{\bf Correlation between the final genome size and the final number of evolved traits.} Black circles represent the combined data from populations with 10, 100, 1000, and 10000 individuals. Only replicates that survived all 2.5x$10^5$ generations were included.}
\label{fig5}
\end{figure}

That the trend in genome size evolution and in phenotypic complexity evolution are mirrored suggests that the evolution of larger genomes enables the evolution of increased phenotypic complexity. To establish a link between the two, we performed two tests. First, we examine the correlation between genome size and phenotypic complexity across all populations. Phenotypic complexity is positively correlated with genome size (Fig.~\ref{fig5}, Spearman's $\rho\approx$0.72; $p<$ 2.3 x $10^{-57}$ ), suggesting that it was increased genome size that allowed for the evolution of increased phenotypic complexity. However, there are two potential mechanisms that could cause an increased genome size to result in increased phenotypic complexity. The first mechanism is that a larger genome has more room for functional content. The second is that a larger genome results in an increased genomic mutation rate and a potentially faster rate of evolution. To examine the role of an increased mutation rate in driving the evolution of phenotypic complexity, we evolved a further one hundred populations of ten individuals with a fixed genome mutation rate of $1.5 \times 10^{-1}$ (i.e., the ancestral genomic mutation rate). Under this condition, no population went extinct (as opposed to forty-seven in the variable mutation rate treatment). The fixed genomic mutation rate populations evolved a median of 2 phenotypic traits compared to the variable genomic mutation rate populations that had evolved a median of 4 phenotypic traits (Fig. S3). These data demonstrate that the increased genomic mutation rate that follows from larger genomes does increase the evolution of phenotypic complexity. However, even with a fixed genomic mutation rate, the smallest populations still evolved a greater median number of traits (on average 2 traits) than every other population size. Thus, while an increased genomic mutation rate (due to increased sequence length) indeed enhances the evolution of phenotypic complexity, small populations still possess an evolutionary advantage due to drift-driven increases in genome size only.

\subsection*{Non-Functional Insertions}
In the previous experiments, large populations evolved larger genomes and greater phenotypic complexity because they fixed rare beneficial insertions. Next, we more closely examine the finding that beneficial insertions are necessary for the evolution of complexity in large populations. We repeated the experiments with the same population sizes and mutation rates, except we changed how insertions worked. Instead of inserting one of the twenty-six instructions that compose the Avida instruction set, we inserted ``blank" instructions into the genome (see Methods for details). These blank instructions cannot be beneficial (on their own or in combination with existing instructions) and would have to be further mutated to lead to the evolution of phenotypic complexity. In this treatment, greater phenotypic complexity in large populations would require a two-step mutational process, as opposed to the single step in a beneficial insertion. 

We saw no qualitative difference in the trend between these experiments and the original experiments (Fig. S4). Very small and large populations still both evolved the largest genomes and the greatest phenotypic complexity. Populations from all population sizes evolved longer genomes and more phenotypic traits in this treatment (Fig. S4) than in the original treatment (Fig.~\ref{fig1} and Fig.~\ref{fig4}). The fraction of fixed insertions that were under positive selection decreased for every population size compared to the original experiments, as expected from the insertion of non-functional instructions (Fig. S5). We observed an increased rate of extinction in the very small populations, with only 2 populations with ten individuals and 25 populations with twenty individuals surviving the experiment. Population extinction was likely enhanced by the increased growth in genome size in these experiments as compared to the original experiments. 

\subsection*{Deletion Bias}
Finally, we performed experiments to test the effect of a deletion bias (a higher fraction of deletions among all indels) alters the relationship between population size and the evolution of complexity. A biased ratio of deletion to insertion mutations is found in biological organisms across the tree of life, especially in bacteria~\cite{mira2001deletional,kuo2009deletional}. In these experiments we set the ratio of deletions to insertions as 9:1, but kept the total indel mutation rate as in the original experiments.  In this treatment, only one population with ten individuals went extinct, as opposed to 47 populations in the original treatment. However, the advantage towards evolving complexity previously enjoyed by small populations vanished (Fig. S6). The median genome size increased as the population size increased for all populations sizes. Only the largest populations evolved a median number of novel phenotypic traits greater than zero. These results suggest that it is not only the role of genetic drift, but the equal frequency of insertions and deletions that results in the increased genome size and phenotypic complexity in small populations. 

\section*{Discussion}

The idea that small populations could have an evolutionary advantage over large populations dates back to Wright and his Shifting Balance theory~\cite{wright1932roles}. More recently, a potential small population advantage has been demonstrated both theoretically~\cite{jain2011evolutionary} and experimentally~\cite{rozen2008heterogeneous}, but only in regard to short-term increases in fitness. The Mutational Burden hypothesis provides an evolutionary mechanism that gives small populations an advantage towards increased phenotypic complexity~\cite{lynch2007frailty,koonin2012logic}. However, an experimental demonstration of this advantage is lacking. Our study provides further insight into the conditions that give small populations such an evolutionary advantage. We confirmed that small populations do evolve larger genomes due to the increased fixation of slightly deleterious mutations, as predicted~\cite{lynch2003origins}. We also showed how small populations have an increased potential to later evolve increased phenotypic complexity in small populations through the larger genomes generated by increased genetic drift~\cite{koonin2004non,lynch2007frailty}. 

Our work also shows that this evolutionary advantage of small populations is limited by an increased rate of population extinction.  Such a trend between the evolution of large genomes and an increased rate of extinction is seen in some multicellular eukaryote clades~\cite{vinogradov2003selfish,vinogradov2004genome}. These small populations are still likely to have a larger risk of extinction beyond that caused by population-genetic risks such as Muller's ratchet~\cite{muller1964relation} and mutational meltdowns~\cite{lynch1993mutational,zeyl2001mutational}. Ecological stressors increase extinction risk~\cite{smith1989causes} and small populations are less able to adapt to detrimental environmental changes~\cite{willi2006limits}. Our results concerning extinction, combined with the risk of other factors not examined here, suggest that the likelihood of a small population using genetic drift to evolve greater complexity without an increased risk of extinction may be limited. However, it is possible that multiple small populations could reduce the risk of extinction without reducing the evolution of complexity; future work should consider the interplay between population size and the evolution of complexity within a metapopulation of small populations. 

Large populations also evolved greater genome sizes and phenotypic complexity. In our original experiments, genome evolution in large populations was driven by the fixation of rare beneficial insertions (Fig.~\ref{fig4}). While it is likely that many gene duplications are not under positive selection and lost due to genetic drift and mutation accumulation~\cite{lynch2000evolutionary}, some, especially those resulting in the amplification of gene expression, can be immediately beneficial and later lead to increased phenotypic complexity~\cite{hughes1994evolution,bergthorsson2007ohno,blount2012genomic,nasvall2012real}. Due to the increased mutation supply, these events would occur at a greater frequency in large populations~\cite{walsh1995often} and possibly lead to an increased probability of the evolution of complexity there. However, we also found that large populations did not require this large supply of beneficial insertions. Even when insertion mutations added non-functional instructions and further point mutations were required to evolve functional traits, large populations still evolved complexity similar to that evolved in small populations. These results suggest that purifying selection may not limit the evolution of complexity in large populations. Finally, we found that when deletions occur at a much greater frequency than insertions, only large populations have an evolutionary advantage towards complexity. As many bacteria do have a bias towards deletions~\cite{daubin2004comment,kuo2009consequences}, this result suggests that large microbial populations can have an evolutionary advantage over small microbial populations for evolving novel traits after all.

Such a trend where both large and small, but not intermediate-sized populations have an evolutionary advantage has already been theoretically proposed elsewhere. Weissman et al.\ showed that both small and large populations cross fitness valleys more easily than intermediate-sized populations~\cite{weissman2009rate}. Small populations valley-crossed due to genetic drift and large populations did so due to an increased supply of double mutants. Ochs and Desai also showed that intermediate-sized populations evolved to a lower fitness peak compared to small or large populations when valley crossing was required for reach a higher peak~\cite{ochs2015competition}. We found similar results, but from different evolutionary mechanisms. Here, populations needed to increase in genome size in order to evolve phenotypic complexity. Additionally, our populations evolved in a complex fitness landscape with many different possible paths to phenotypic complexity. While small populations did fix deleterious insertions to increase genome size, large populations evolved on a different path, either through beneficial insertions (Fig.~\ref{fig3}) or neutral insertions (Fig. S4). It is possible that even larger populations than those evolved here would fix more deleterious insertions, as the likelihood of a further, beneficial mutation arising on the background of a segregating deleterious mutation increases as population size increases. However, our results emphasize that large populations may not be dependent on valley-crossing in some fitness landscapes if alternative evolutionary trajectories exist, even if these trajectories are rare. While the first maps of fitness landscapes suggested mutational paths are small in number~\cite{weinreich2006darwinian}, more recent work suggests that many indirect evolutionary trajectories exist in larger fitness landscapes~\cite{palmer2015delayed}.

Here, we studied the evolution of complexity in haploid asexual digital organisms with an ancestral minimal genome on a frequency-independent fitness landscape. While it is beyond the scope of this work, it is worth considering how adjusting these genotype characteristics would alter our results. It is likely that the ancestral minimal genomes are a requirement for small populations to evolve the same number of novel traits as large populations. If the ancestor organism had a significant amount of non-functional genome content, the mutation supply advantage that large populations have should result in an accelerated rate of phenotypic evolution in large populations~\cite{elena2007effects}. The organisms used here, as in all Avida experiments, are haploid. It is possible that polyploidy would alter the results found here.  However, the implementation of a ploidy cycle in Avida is non-trivial due to the mechanistic style of replication, and so presently other experimental systems would have to be used to explore the role of ploidy in the evolution of phenotypic complexity. 

It is unclear how sexual, instead of asexual, reproduction would change the results. While sexual reproduction can enhance adaptation by combining beneficial mutations that arise in different background, it can also break up beneficial combinations of mutations~\cite{otto2009evolutionary}. One result that may be altered by sexual reproduction is the rate of extinction in small populations, as sex has been found to reduce the rate of mutational meltdowns~\cite{lynch1995mutational}. Weissman et al.\ also demonstrate that the large population advantage towards valley crossing does not exist under high recombination rates~\cite{weissman2010rate}. Sexual reproduction has previously been studied using Avida, but it is more akin to homologous recombination in bacteria~\cite{misevic2006sexual} (as there is no ploidy cycle). Future work should address the role of sexual recombination on the results shown here. Finally, the experiments performed here had no frequency-dependent fitness effects. Previous Avida studies showed that frequency-dependent interactions enhanced the evolution of complexity for a given population size~\cite{walker2012evolutionary,zaman2014coevolution}. It is worth exploring how the presence of frequency-dependent selection alters the evolution of complexity, especially in small populations. The benefits of the diversity seen in frequency-dependent fitness landscapes may be reduced in small populations. The extensions to the experiments performed here would provide a more complete understanding of the role of adaptive and non-adaptive evolutionary processes in the origins of complexity.

\section*{Methods}
\subsection*{Avida}
In order to experimentally test the role of population size and genetic drift in the evolution of complexity, we used the digital evolution system Avida version 2.14~\cite{ofria2009avida}. In Avida, self-replicating computer programs (avidians) compete in a population for a limited supply of CPU (Central Processing Unit) time needed to successfully reproduce. Each avidian consists of a circular haploid genome of computer instructions. During its lifespan, an avidian executes the instructions that compose its genome. After executing certain instructions, it begins to copy its genome. This new copy will eventually be divided off from its mother (reproduction in most Avida experiments is asexual). Because an avidian passes on its genome to its descendants, there is heredity in Avida. As an avidian copies its genome, mutations may occur, resulting in imperfect transmission of hereditary information. This error-prone replication introduces variation into Avida populations. Finally, avidians that differ in instructions (their genetic code) also likely differ in their ability to self-replicate; this results in differential fitness. Therefore, because there is differential fitness, variation, and heredity, an Avida population undergoes evolution by natural selection~\cite{pennock2007models}. This allows researchers to perform experimental evolution in Avida as in microbial systems~\cite{kawecki2012experimental,hindre2012new}. Avida has been successfully used as a model system to explore many topics concerning the evolution of complexity~\cite{lenski1999genome,adami2000evolution,lenski2003evolutionary, goldsby2014evolutionary,zaman2014coevolution}.      

The Avida world consists of a toroidal grid of $N$ cells, where $N$ is the maximum population size. When an avidian successfully divides, its offspring is placed into a cell in the population. While the default setting places the offspring into one of nine neighboring cells of the parent, here the offspring is placed into any cell in the entire population. This simulates a well-mixed environment without spatial structure. When there are empty cells in the population, new offspring are preferentially placed in an empty cell. However, if the population is at its carrying capacity, the individual who is currently occupying the selected cell is replaced by the new offspring (a new individual can also eliminate its parent if that cell is selected). This adds an element of genetic drift into the population as the individual to be removed is selected without regard to fitness. A population can also decrease in size by the death of individuals. An avidian will die without producing offspring if it executes $20L$ instructions without successfully undergoing division, where $L$ is the avidian's genome size. In very small populations, this can lead to population extinction.  

Time in Avida is divided into updates, not generations. This method of time was implemented in order to allow individuals to execute their genomes in parallel. During one update, a set number of instructions are executed across the entire population. The ability to execute one instruction is referred to as a single instruction processing (SIP) unit, and is the CPU ``energy" avidians need to replicate. By default, there are $30N$ SIPs available to the entire population per update, where $N$ is the population size. SIPs are distributed among the individual genotypes within a population in proportion to the trait or traits displayed by an individual. The total amount of SIPs garnered by an individual from traits is called the ``merit". In a homogeneous population of one genotype (clones), where each individual has the same merit, each individual will obtain approximately 30 SIPs per update. However, in a heterogeneous population where merit differs between individuals,  SIPs will be distributed in an uneven manner. That way, individuals with a greater merit will execute and/or replicate a larger proportion of their genome per update and replicate faster, thus having a greater fitness. This places a strong selection pressure on evolving a greater merit. One generation has passed when the population has produced $N$ offspring. Typically (depending on the complexity of an avidian) between 5 and 10 updates pass in one generation. 

A genotype's {\em merit} is increased through the evolution of certain phenotypic traits that form a ``digital metabolism"~\cite{adami2006digital}. These phenotypic traits are the ability (or lack there of) to perform certain Boolean logic calculations on random binary numbers that the environment provides. To do this, an avidian must have the ``genes" to do this--in this case, the right sequence of instructions. First, during an avidian's lifespan, instructions that allow for the input and output of these random binary numbers must be executed. Further instructions should  manipulate those numbers so as to perform the rewarded computations. When a number is then written to the output, the Avida program checks to see whether a logic operation was successfully performed. If so, the the individual that performed the computation consumes a resource tied to the performance of that trait (there are many different codes, that is, combinations of instructions, that will trigger the reward). Resource consumption causes the offspring of that individual to have their merit modified by a factor set by the experimenter. Here, we use the ``Logic-9" environment to reward the performance of nine one- and two-input logic functions~\cite{lenski2003evolutionary}; see Table S1 for the names and specific rewards of each function). Each individual only gains a benefit from performing each function once per generation. There is an infinite supply of resources for the performance of each logic function in the present experiments, making fitness frequency-independent. Because the performance of these logic functions increases merit, they also increase fitness and are under strong positive selection. 

While increases in an individual's merit increase replication speed and thus the individual's fitness, fitness in Avida is implicit and not directly calculated. Unlike simulations of evolutionary dynamics, a genotype's fitness is not set \emph{a priori} by the experimenter. The only way to measure the fitness of an avidian is to run it through its lifecycle and examine its phenotype. This is similar in principle to how bacterial fitness cannot be calculated by examining an individual bacterium's genome, but must be measured through a number of different experiments, such as  competition assays~\cite{lenski1991long}. A genotype's fitness is determined by how many offspring it can produce per unit time. Genotypes that can reproduce faster will out-compete other genotypes, all else being equal. Therefore, evolution will increase a population's fitness through two means. The first is that the population will evolve individuals with a greater number of phenotypic traits and thus with a greater merit, as explained above. The second way to increase replication speed is by optimizing (shortening) the replication time. This occurs either by shrinking the genome, which results in fewer instructions that need to be copied and replicated, or by optimizing genome architecture for faster replication. Fitness $w$ in Avida is estimated by the following equation:   
\begin{equation}\label{eq:fitness} w \approx \frac{\rm merit}{\rm replication~time}
\end{equation}

For an avidian to be able to successfully reproduce, it must first allocate memory for the new individual, copy its genome into the allocated memory space, and then divide off the daughter organism. As instructions are copied, the avidian may inaccurately copy some instructions into the newly allocated memory at a rate set by the experimenter. Additionally, upon division, insertions and deletions of a single instructions occur at (possibly different) rates set by the experimenter. Finally, larger insertions or deletions (indels) can occur when an avidian divides into two daughter genomes if the division occurs unevenly. In most cases, this results in the creation of one larger and one smaller genome and both of these are non-viable. However, in rare cases, one of these new genotypes is able to reproduce, resulting in a large change in genome size in that individual's descendants. Because this mutation through inaccurate division is a characteristic of a genome and thus emergent, the rate at which it occurs is not set by the experimenter. 

\subsection*{Experimental Design}
We used four experimental designs (treatments) to explore how population size determines the evolution of complexity: the original experiments, the non-functional insertion experiments, the fixed genomic mutation rate experiments, and the deletion bias experiments. For all experiments, we evolved populations of size $N$=\{10,100,1000,10000\} for $2.5\times10^5$ generations under 100-fold replication. For the original treatment, we also performed experiments with population sizes of $N$=\{20,30,40,50,60,70,80,90\}. All populations were initiated at full size $N$ with an altered version of the standard Avida start organism~\cite{ofria2009avida}. The alteration was the removal of all non-essential genome content (85 {\tt nop-c} instructions). This reduced the genome size of the ancestor organisms from 100 instructions to only 15 instructions.

For the original experiments, point mutations occurred at a rate of 0.01 mutations per instruction copied, and insertions and deletions at 0.005 events per division. Insertions and deletions occur at most once per division. The ancestor thus started with a genomic mutation rate of 0.15 mutations per generation (0.01 mutations/instruction copied $\times$ fifteen instructions copied per generation), but this changes over the course of the experiment as genome size evolves. These experiments are similar to most standard Avida experiments, with the exception of a smaller genome size (fifteen instructions) for the ancestral organism. 

For the remainder of the experimental settings, one of the above settings was changed to examine a specific effect. For the experiments where the genomic mutation rate was fixed, point mutations occurred at a rate of 0.15 mutations {\em per division}, independently of genome size. This fixed the mutation rate at 0.15 mutations/genome/generation. For the non-functional insertion experiments, the mutation rates were the same as in the original experiments. However, instead of inserting one of the twenty-six instructions from the Avida instruction set (see~\cite{ofria2009avida} for the Avida instruction set), ``blank" instructions called {\tt nop-x} were inserted. These instructions had no function and would usually have no effect when executed by the Avidian. Finally, for the deletion bias experiments, point mutations occurred at the same rate as in the standard experiments. However, insertions and deletions did not occur at the same rate. Insertions occurred at a rate of 0.001 per division and deletions occurred at a rate of 0.009 per division. This kept the total mutation rate equal to the other experimental treatments, while altering the ratio of insertions to deletions.

\subsection*{Data Analysis}
In order to analyze the evolution of complexity in each population, we extracted the individual with the greatest fitness at the end of each experiment (the ``dominant" type). We then calculated relevant statistics for each of these genotypes by running them through Avida's {\em analyze mode}. This mode allows us to run each genotype through its lifecycle in isolation, and calculate its fitness, its genome size, whether it performs any logic functions, and whether it produces viable offspring, among other characteristics. To measure the evolution of phenotypic complexity, we determined how many unique logic calculations each genotype could perform. This is a similar calculation in concept to a measure of phenotypic complexity used previously~\cite{tenaillon2007quantifying} in population genetics.

To examine why certain population sizes evolved larger genomes, we examined the ``line of descent" (LOD) of the dominant type~\cite{lenski2003evolutionary}. An LOD contains every intermediate genotype between the final dominant individual and the ancestral genotype that initialized each population. This line provides a perfect fossil record to examine all of the mutations, insertions, and deletions that led to the final dominant genotype for each population. We also calculated the selection coefficient $s$ for each mutation, defined as the ratio of the offspring's fitness to the parent's fitness minus one. We defined beneficial mutations as those with $s>0$ and deleterious mutations as those with $s<0$ (this ignores classifying slightly beneficial and slightly deleterious mutations as neutral.) We determined the number of beneficial insertion mutations by counting those insertions on the LOD with $s>\frac{1}{N}$, where $N$ is the population size. These are beneficial mutations that are not nearly-neutral and hence should be under positive selection. Using  $s>\frac{1}{N}$ is only an approximation, as the equation for a nearly neutral mutation is $|s|\ll\frac{1}{N_e}$, where $N_e$ is the effective population size~\cite{ohta1996development}.
We also examined those mutations that had a slightly deleterious effect on fitness, i.e., those whose selection coefficient was $-\frac{1}{N}<s<0$.

\section*{Acknowledgments}
We thank members of the Adami lab for discussions. This work was supported in part by Michigan State University through computational resources provided by the Institute for Cyber-Enabled Research.
This work was supported in part by Michigan State University through computational resources provided by the Institute for Cyber-Enabled Research, by a Michigan State University Distinguished Fellowship to TL, and by the National Science Foundation's BEACON Center for the Study of Evolution in Action, under contract No. DBI-0939454.

\section*{Supporting Information}
\newpage
\begin{table*}
\begin{center}
\caption{{\bf Merit rewards for the evolution of phenotypic traits.}}
  \begin{tabular}{ | c | c |}
    \hline
    Boolean Logic Calculaton & Merit Multiplier \\ \hline
    NOT & 2 \\ \hline
    NAND & 2 \\ \hline
    ORNOT & 4 \\ \hline
    AND & 4 \\ \hline
    ANDNOT & 8 \\ \hline
    OR & 8 \\ \hline
    NOR & 16 \\ \hline
    XOR & 16 \\ \hline
    XNOR (Equals) & 32 \\ \hline
  \end{tabular}
\end{center}
\end{table*}
\label{S1 Table}

\makeatletter 
\renewcommand{\thefigure}{S\@arabic\c@figure}
\makeatother
\setcounter{figure}{0}

\begin{figure}
\begin{center} 
\includegraphics[height=3in,width=4.5in]{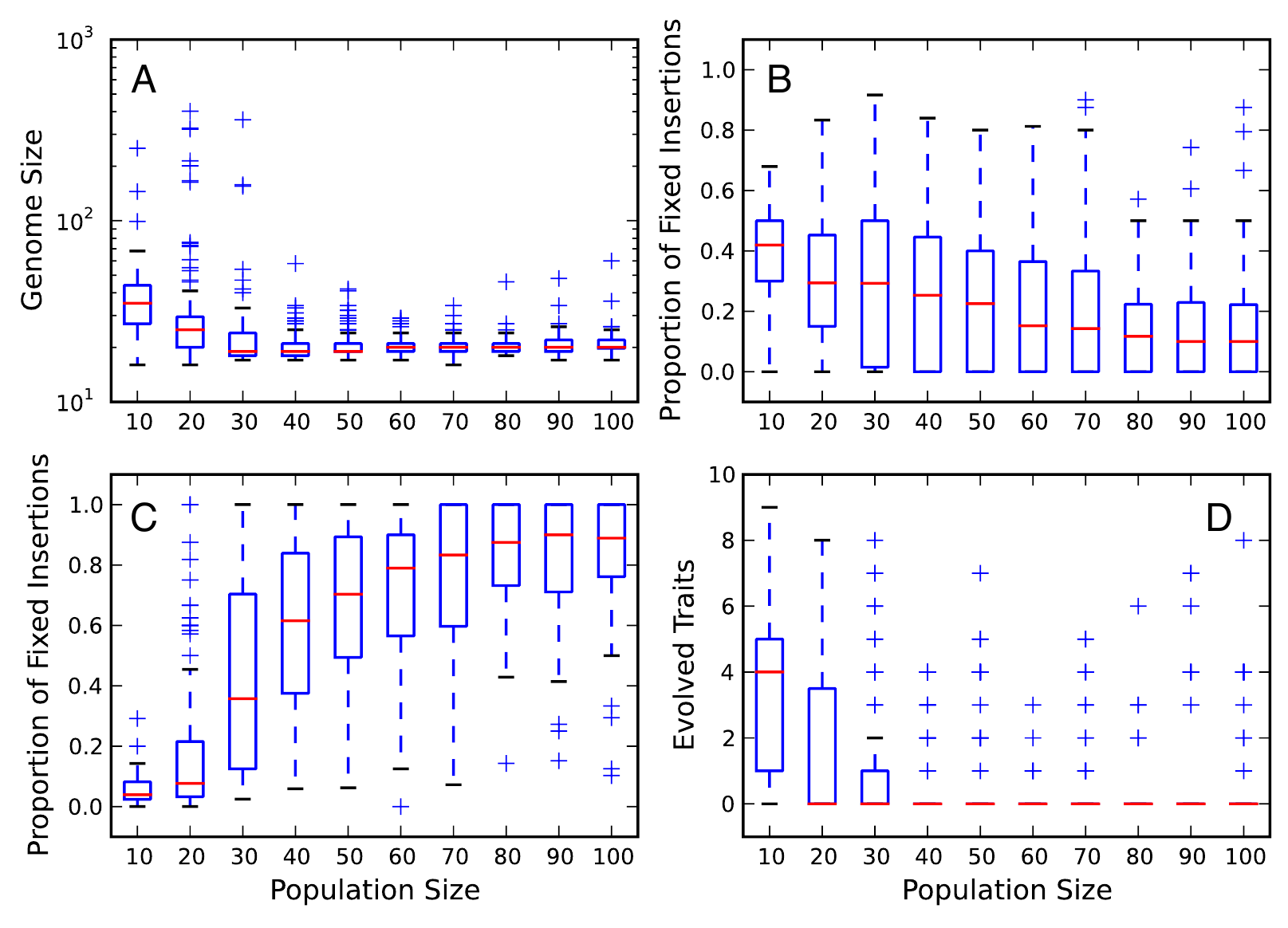} 
\caption{{\bf Evolution of complexity for small population sizes.} Statistics showed in the main text for population sizes ranging from 10 to 100 individuals. Data for populations with 10 and 100 individuals are the same as in the main text. A: Evolution of genome size. B: Proportion of fixed insertions that were slightly-deleterious. C: Proportion of fixed insertions that were under positive selection. D: Number of evolved novel phenotypic traits. Red lines are the median values for each population size. The upper and lower limits of each box denote the third and first quartile, respectively. Whiskers are 1.5 times the relevant quartile value. Plus signs denote those data points beyond the whiskers. Data represent only those populations that did not go extinct.}
\end{center}
\end{figure}

\begin{figure}
\begin{center} 
\includegraphics[height=3in,width=4.5in]{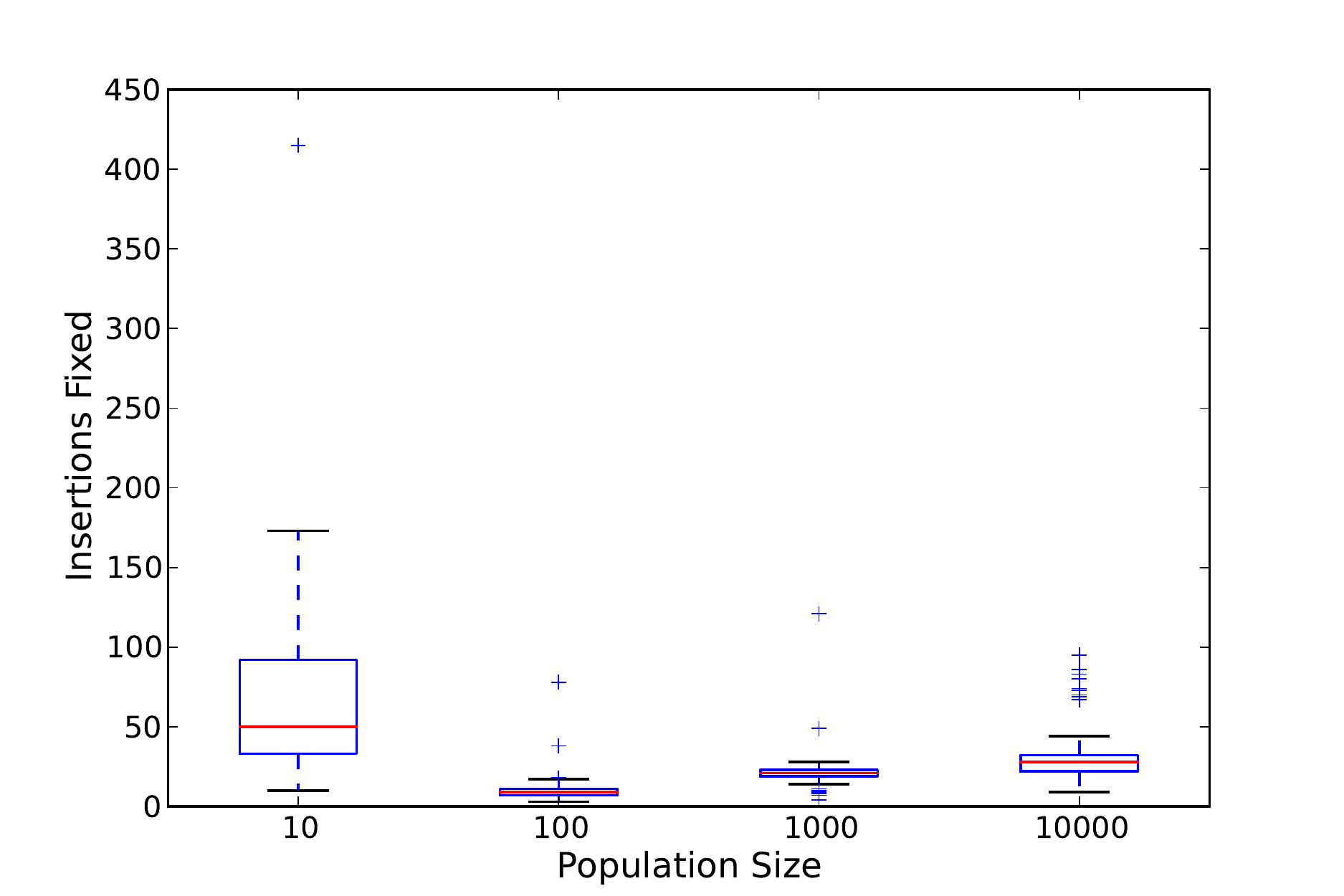} 
\caption{{\bf The number of insertions fixed as a function of population size.} Red lines are the median values for each population size. The upper and lower limits of each box denote the third and first quartile, respectively. Whiskers are 1.5 times the relevant quartile value. Plus signs denote those data points beyond the whiskers. Data represent only those populations that did not go extinct.}
\end{center}
\end{figure}

\begin{figure}
\begin{center} 
\includegraphics[height=3in,width=4.5in]{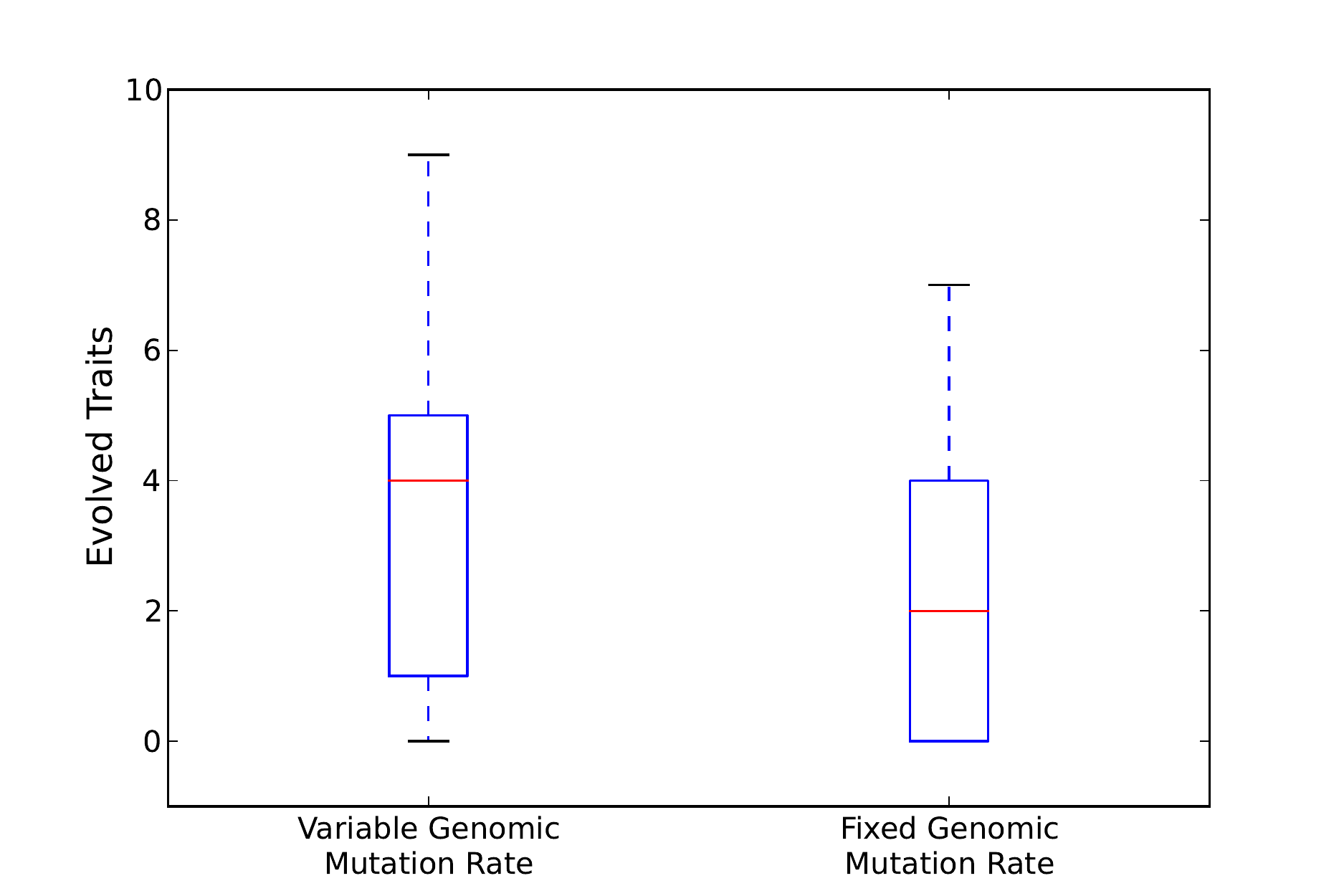} 
\caption{{\bf The effect of a fixed mutation rate on the evolution of phenotypic complexity.} The variable genomic mutation rate treatment represents the data from when the genomic point mutation rate is $10^{-1} \times L$, were $L$ is the genome size. The fixed genomic mutation rate treatment represents the data from when the genomic point mutation rate was fixed at $1.5\times 10^{-1}$, independent of the genome size. Red lines are the median values for each population size. The upper and lower limits of each box denote the third and first quartile, respectively. Whiskers are 1.5 times the relevant quartile value. Plus signs denote those data points beyond the whiskers. Data represent only those populations that did not go extinct.}
\end{center}
\end{figure}

\begin{figure}
\begin{center} 
\includegraphics[height=3in,width=4.5in]{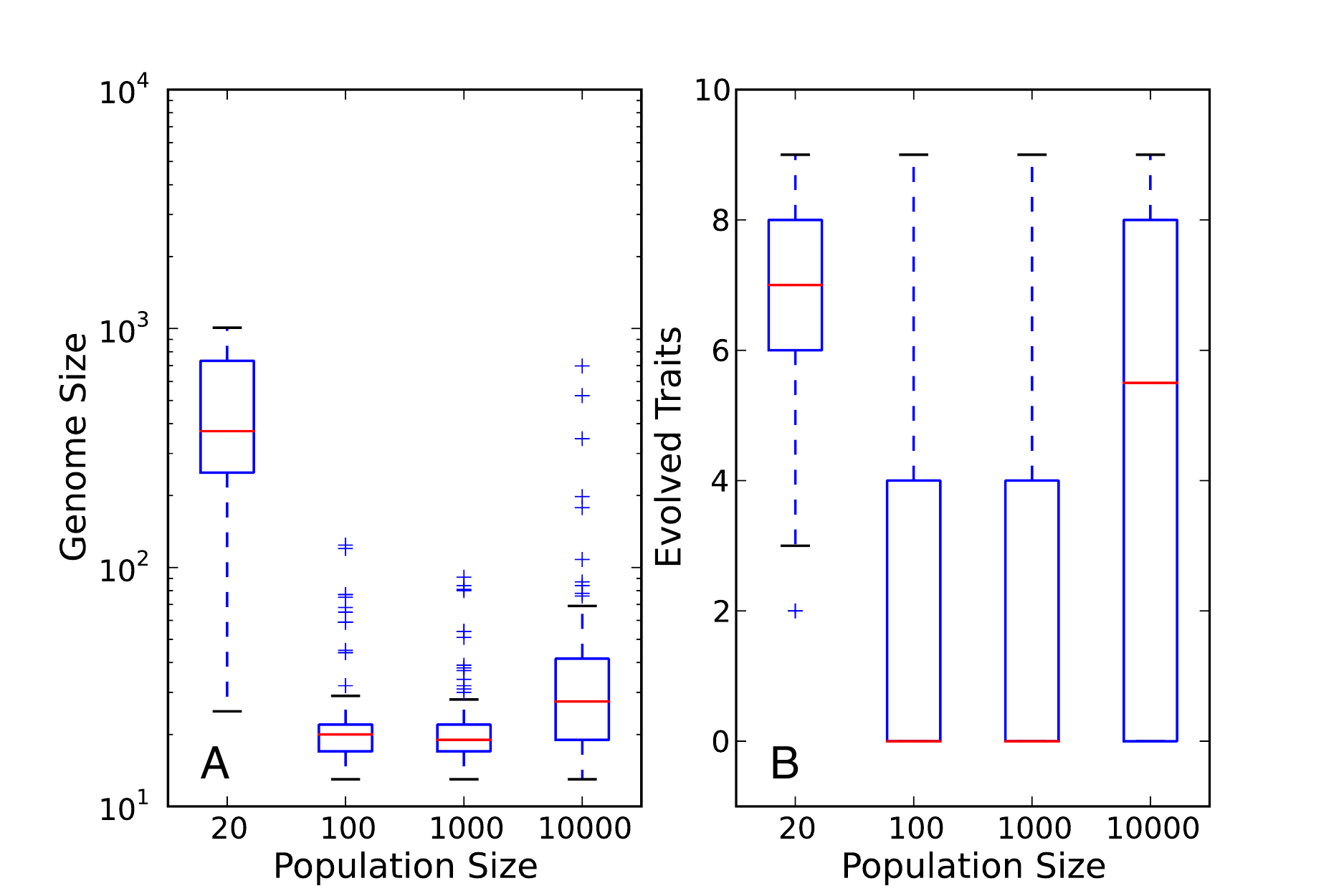} 
\caption{{\bf The evolution of complexity in the non-functional insertion treatment.} All subplots are a function of the population size. A: The final genome size. B: The final number of evolved phenotypic traits. Populations with twenty individuals are shown instead of those with ten individuals due to the high extinction rates of populations with ten individuals. Red lines are the median values for each population size. The upper and lower limits of each box denote the third and first quartile, respectively. Whiskers are 1.5 times the relevant quartile value. Plus signs denote those data points beyond the whiskers. Data represent only those populations that did not go extinct.}
\end{center}
\end{figure}

\begin{figure}
\begin{center} 
\includegraphics[width=4.5in]{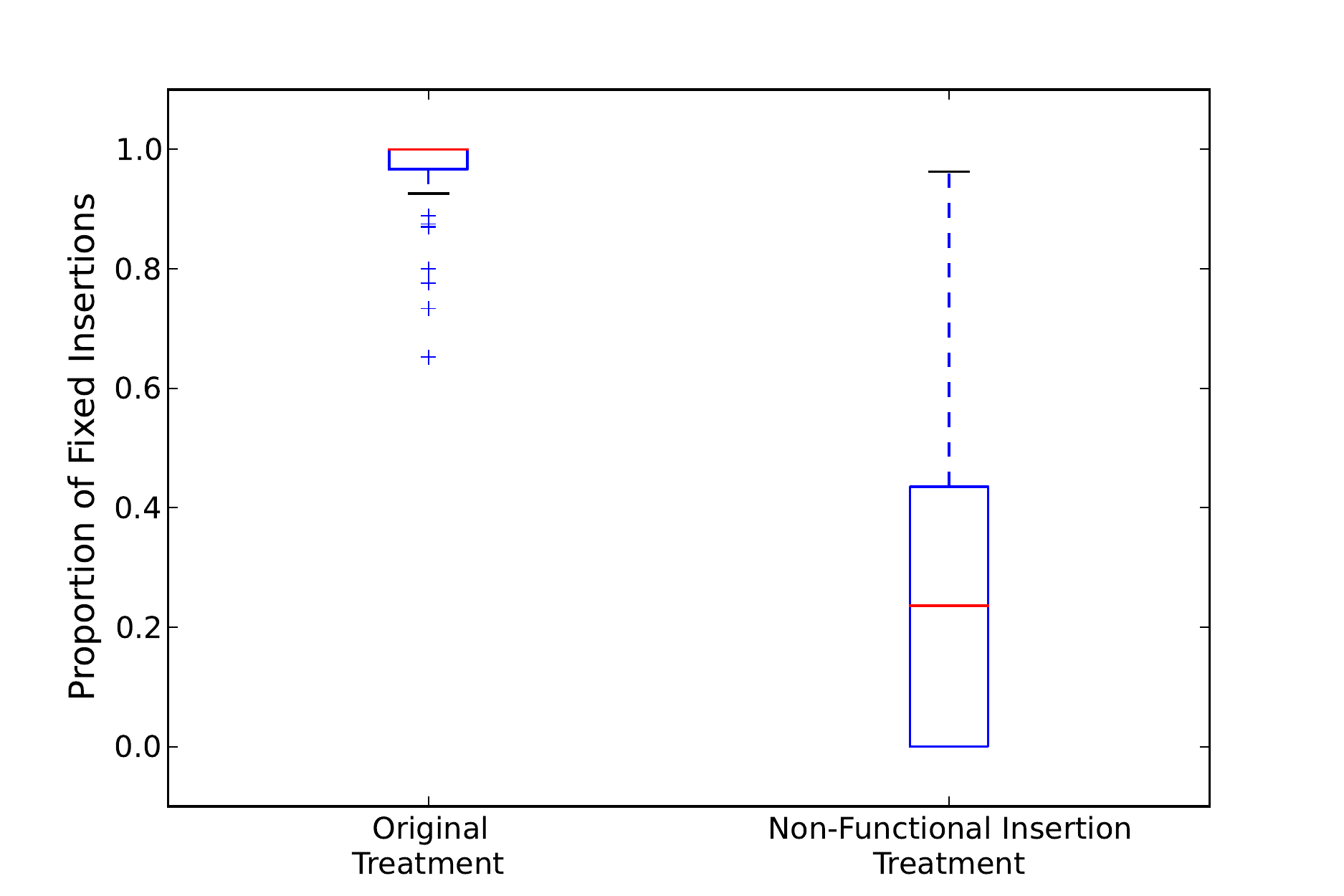} 
\caption{{\bf The proportion of fixed insertions that were under positive selection in the non-functional insertion treatment compared to the original treatment for populations with $10^4$ individuals.} Red lines are the median values for each population size. The upper and lower limits of each box denote the third and first quartile, respectively. Whiskers are 1.5 times the relevant quartile value. Plus signs denote those data points beyond the whiskers. Data represent only those populations that did not go extinct.}
\end{center}
\end{figure}

\begin{figure}
\begin{center} 
\includegraphics[height=3in,width=4.5in]{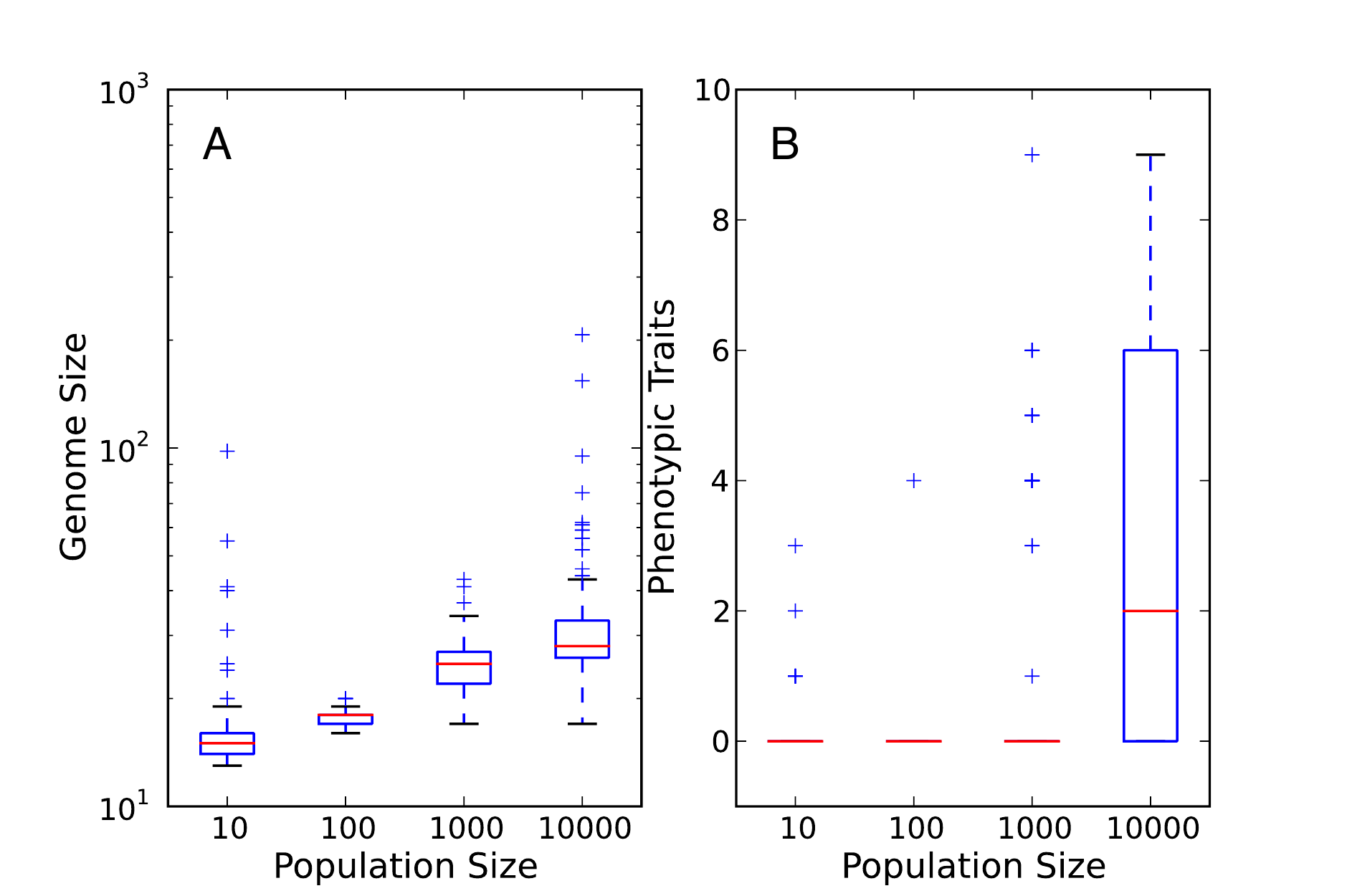} 
\caption{{\bf The evolution of complexity in the deletion bias treatment.} All subplots are a function of the population size. A: The final genome size. B: The final number of evolved phenotypic traits. Red lines are the median values for each population size. The upper and lower limits of each box denote the third and first quartile, respectively. Whiskers are 1.5 times the relevant quartile value. Plus signs denote those data points beyond the whiskers. Data represent only those populations that did not go extinct.}
\end{center}
\end{figure}

\clearpage


\newpage
\begin{thebibliography}{10}

\bibitem{bonner1988evolution}
Bonner JT.
\newblock The evolution of complexity by means of natural selection.
\newblock Princeton: Princeton University Press; 1988.

\bibitem{adami2000evolution}
Adami C, Ofria C, Collier TC.
\newblock Evolution of biological complexity.
\newblock Proceedings of the National Academy of Sciences. 2000;97:4463--4468.

\bibitem{koonin2004non}
Koonin EV.
\newblock A non-adaptationist perspective on evolution of genomic complexity or
  the continued dethroning of man.
\newblock Cell Cycle. 2004;3:278--283.

\bibitem{lynch2007frailty}
Lynch M.
\newblock The frailty of adaptive hypotheses for the origins of organismal
  complexity.
\newblock Proceedings of the National Academy of Sciences. 2007;104(Suppl
  1):8597--8604.

\bibitem{tenaillon2007quantifying}
Tenaillon O, Silander OK, Uzan JP, Chao L.
\newblock Quantifying organismal complexity using a population genetic
  approach.
\newblock PLoS One. 2007;2:e217.

\bibitem{mcshea2010biology}
McShea DW, Brandon RN.
\newblock Biology's first law: the tendency for diversity and complexity to
  increase in evolutionary systems.
\newblock Chicago: University of Chicago Press; 2010.

\bibitem{kimura1984neutral}
Kimura M.
\newblock The neutral theory of molecular evolution.
\newblock New York: Cambridge University Press; 1984.

\bibitem{ohta1992nearly}
Ohta T.
\newblock The nearly neutral theory of molecular evolution.
\newblock Annual Review of Ecology and Systematics. 1992;23:263--286.

\bibitem{gillespie2000genetic}
Gillespie JH.
\newblock Genetic drift in an infinite population: the pseudohitchhiking model.
\newblock Genetics. 2000;155:909--919.

\bibitem{lynch2011repatterning}
Lynch M, Bobay LM, Catania F, Gout JF, Rho M.
\newblock The repatterning of eukaryotic genomes by random genetic drift.
\newblock Annual Review of Genomics and Human Genetics. 2011;12:347--366.

\bibitem{eddy2012c}
Eddy SR.
\newblock The C-value paradox, junk DNA and ENCODE.
\newblock Current Biology. 2012;22:R898--R899.

\bibitem{palazzo2014case}
Palazzo AF, Gregory TR.
\newblock The case for junk DNA.
\newblock PLoS Genet. 2014;10:e1004351.

\bibitem{travisano1995experimental}
Travisano M, Mongold JA, Bennett AF, Lenski RE.
\newblock Experimental tests of the roles of adaptation, chance, and history in
  evolution.
\newblock Science. 1995;267:87--90.

\bibitem{wagenaar2004influence}
Wagenaar DA, Adami C.
\newblock Influence of chance, history, and adaptation on digital evolution.
\newblock Artificial Life. 2004;10:181--190.

\bibitem{lachapelle2015repeatability}
Lachapelle J, Reid J, Colegrave N.
\newblock Repeatability of adaptation in experimental populations of different
  sizes.
\newblock Proceedings of the Royal Society of London B: Biological Sciences.
  2015;282:20143033.

\bibitem{blount2008historical}
Blount ZD, Borland CZ, Lenski RE.
\newblock Historical contingency and the evolution of a key innovation in an
  experimental population of Escherichia coli.
\newblock Proceedings of the National Academy of Sciences.
  2008;105(23):7899--7906.

\bibitem{meyer2012repeatability}
Meyer JR, Dobias DT, Weitz JS, Barrick JE, Quick RT, Lenski RE.
\newblock Repeatability and contingency in the evolution of a key innovation in
  phage lambda.
\newblock Science. 2012;335:428--432.

\bibitem{wagner2011origins}
Wagner A.
\newblock The origins of evolutionary innovations: a theory of transformative
  change in living systems.
\newblock New York: Oxford University Press; 2011.

\bibitem{kawecki2012experimental}
Kawecki TJ, Lenski RE, Ebert D, Hollis B, Olivieri I, Whitlock MC.
\newblock Experimental evolution.
\newblock Trends in Ecology \& Evolution. 2012;27:547--560.

\bibitem{masel2011genetic}
Masel J.
\newblock Genetic drift.
\newblock Current Biology. 2011;21:R837--R838.

\bibitem{schoustra2009properties}
Schoustra SE, Bataillon T, Gifford DR, Kassen R, et~al.
\newblock The properties of adaptive walks in evolving populations of fungus.
\newblock PLoS biology. 2009;7:2474.

\bibitem{quandt2015fine}
Quandt EM, Gollihar J, Blount ZD, Ellington AD, Georgiou G, Barrick JE.
\newblock Fine-tuning citrate synthase flux potentiates and refines metabolic
  innovation in the Lenski evolution experiment.
\newblock eLife. 2015;4:e09696.

\bibitem{burmeister2015selection}
Burmeister A, Lenski R, Meyer J. Selection for intermediate genotypes enables a
  key innovation in phage lambda; 2015.
\newblock bioRxiv http://dx.doi.org/10.1101/018606.

\bibitem{wagner2008neutralism}
Wagner A.
\newblock Neutralism and selectionism: a network-based reconciliation.
\newblock Nature Reviews Genetics. 2008;9:965--974.

\bibitem{wright1932roles}
Wright S.
\newblock The roles of mutation, inbreeding, crossbreeding, and selection in
  evolution.
\newblock In: Proceedings of the Sixth International Congress of Genetics.
  vol.~1; 1932. p. 356--366.

\bibitem{rozen2008heterogeneous}
Rozen DE, Habets MGJL, Handel A, De~Visser JAGM.
\newblock Heterogeneous adaptive trajectories of small populations on complex
  fitness landscapes.
\newblock PLoS One. 2008;3:e1715--e1715.

\bibitem{jain2011evolutionary}
Jain K, Krug J, Park SC.
\newblock Evolutionary advantage of small populations on complex fitness
  landscapes.
\newblock Evolution. 2011;65:1945--1955.

\bibitem{lynch2003origins}
Lynch M, Conery JS.
\newblock The origins of genome complexity.
\newblock Science. 2003;302:1401--1404.

\bibitem{lynch2007origins}
Lynch M.
\newblock The origins of genome architecture.
\newblock Sunderland: Sinauer Associates; 2007.

\bibitem{lynch2006streamlining}
Lynch M.
\newblock Streamlining and simplification of microbial genome architecture.
\newblock Annu Rev Microbiol. 2006;60:327--349.

\bibitem{batut2014reductive}
Batut B, Knibbe C, Marais G, Daubin V.
\newblock Reductive genome evolution at both ends of the bacterial population
  size spectrum.
\newblock Nature Reviews Microbiology. 2014;12:841--850.

\bibitem{zwart2014experimental}
Zwart MP, Willemsen A, Dar{\`o}s JA, Elena SF.
\newblock Experimental evolution of pseudogenization and gene loss in a plant
  RNA virus.
\newblock Molecular Biology and Evolution. 2014;31:121--134.

\bibitem{koonin2012logic}
Koonin EV.
\newblock The logic of chance: the nature and origin of biological evolution.
\newblock Upper Saddle River: FT press; 2012.

\bibitem{weissman2009rate}
Weissman DB, Desai MM, Fisher DS, Feldman MW.
\newblock The rate at which asexual populations cross fitness valleys.
\newblock Theoretical Population Biology. 2009;75:286--300.

\bibitem{weissman2010rate}
Weissman DB, Feldman MW, Fisher DS.
\newblock The rate of fitness-valley crossing in sexual populations.
\newblock Genetics. 2010;186:1389--1410.

\bibitem{ochs2015competition}
Ochs IE, Desai MM.
\newblock The competition between simple and complex evolutionary trajectories
  in asexual populations.
\newblock BMC Evolutionary Biology. 2015;15:55.

\bibitem{adami2006digital}
Adami C.
\newblock Digital genetics: Unravelling the genetic basis of evolution.
\newblock Nature Reviews Genetics. 2006;7:109--118.

\bibitem{yedid2002macroevolution}
Yedid G, Bell G.
\newblock Macroevolution simulated with autonomously replicating computer
  programs.
\newblock Nature. 2002;420:810--812.

\bibitem{bell2016experimental}
Bell G; The~Royal Society.
\newblock Experimental macroevolution.
\newblock Proc R Soc B. 2016;283:20152547.

\bibitem{batut2013silico}
Batut B, Parsons DP, Fischer S, Beslon G, Knibbe C.
\newblock In silico experimental evolution: A tool to test evolutionary
  scenarios.
\newblock BMC bioinformatics. 2013;14:1.

\bibitem{elena2003evolution}
Elena SF, Lenski RE.
\newblock Evolution experiments with microorganisms: The dynamics and genetic
  bases of adaptation.
\newblock Nature Reviews Genetics. 2003;4:457--469.

\bibitem{ofria2009avida}
Ofria C, Bryson DM, Wilke CO.
\newblock Avida: A Software Platform for Research in Computational Evolutionary
  Biology.
\newblock In: Maciej~Komosinski AA, editor. Artificial Life Models in Software.
  Springer London; 2009. p. 3--35.

\bibitem{Adami1998}
Adami C.
\newblock Introduction to Artificial Life.
\newblock New York: Springer Verlag; 1998.

\bibitem{WilkeAdami2002}
Wilke CO, Adami C.
\newblock {The biology of digital organisms}.
\newblock Trends in Ecology \& Evolution. 2002;17:528--532.

\bibitem{mira2001deletional}
Mira A, Ochman H, Moran NA.
\newblock Deletional bias and the evolution of bacterial genomes.
\newblock Trends in Genetics. 2001;17:589--596.

\bibitem{kuo2009deletional}
Kuo CH, Ochman H.
\newblock Deletional bias across the three domains of life.
\newblock Genome Biology and Evolution. 2009;1:145--152.

\bibitem{vinogradov2003selfish}
Vinogradov AE.
\newblock Selfish DNA is maladaptive: Evidence from the plant Red List.
\newblock Trends in Genetics. 2003;19:609--614.

\bibitem{vinogradov2004genome}
Vinogradov AE.
\newblock Genome size and extinction risk in vertebrates.
\newblock Proceedings of the Royal Society of London B: Biological Sciences.
  2004;271:1701--1706.

\bibitem{muller1964relation}
Muller HJ.
\newblock The relation of recombination to mutational advance.
\newblock Mutation Research. 1964;1:2--9.

\bibitem{lynch1993mutational}
Lynch M, B{\"u}rger R, Butcher D, Gabriel W.
\newblock The mutational meltdown in asexual populations.
\newblock Journal of Heredity. 1993;84:339--344.

\bibitem{zeyl2001mutational}
Zeyl C, Mizesko M, De~Visser J.
\newblock Mutational meltdown in laboratory yeast populations.
\newblock Evolution. 2001;55:909--917.

\bibitem{smith1989causes}
Maynard~Smith J.
\newblock The causes of extinction.
\newblock Philosophical Transactions of the Royal Society B. 1989;325:241--252.

\bibitem{willi2006limits}
Willi Y, Van~Buskirk J, Hoffmann AA.
\newblock Limits to the adaptive potential of small populations.
\newblock Annual Review of Ecology, Evolution, and Systematics.
  2006;37:433--458.

\bibitem{lynch2000evolutionary}
Lynch M, Conery JS.
\newblock The evolutionary fate and consequences of duplicate genes.
\newblock Science. 2000;290:1151--1155.

\bibitem{hughes1994evolution}
Hughes AL.
\newblock The evolution of functionally novel proteins after gene duplication.
\newblock Proceedings of the Royal Society of London B. 1994;256:119--124.

\bibitem{bergthorsson2007ohno}
Bergthorsson U, Andersson DI, Roth JR.
\newblock Ohno's dilemma: Evolution of new genes under continuous selection.
\newblock Proceedings of the National Academy of Sciences.
  2007;104:17004--17009.

\bibitem{blount2012genomic}
Blount ZD, Barrick JE, Davidson CJ, Lenski RE.
\newblock Genomic analysis of a key innovation in an experimental Escherichia
  coli population.
\newblock Nature. 2012;489:513--518.

\bibitem{nasvall2012real}
N{\"a}svall J, Sun L, Roth JR, Andersson DI.
\newblock Real-time evolution of new genes by innovation, amplification, and
  divergence.
\newblock Science. 2012;338:384--387.

\bibitem{walsh1995often}
Walsh JB.
\newblock How often do duplicated genes evolve new functions?
\newblock Genetics. 1995;139:421--428.

\bibitem{daubin2004comment}
Daubin V, Moran NA.
\newblock Comment on "The origins of genome complexity".
\newblock Science. 2004;306:978--978.

\bibitem{kuo2009consequences}
Kuo CH, Moran NA, Ochman H.
\newblock The consequences of genetic drift for bacterial genome complexity.
\newblock Genome Research. 2009;19:1450--1454.

\bibitem{weinreich2006darwinian}
Weinreich DM, Delaney NF, DePristo MA, Hartl DL.
\newblock Darwinian evolution can follow only very few mutational paths to
  fitter proteins.
\newblock Science. 2006;312:111--114.

\bibitem{palmer2015delayed}
Palmer AC, Toprak E, Baym M, Kim S, Veres A, Bershtein S, et~al.
\newblock Delayed commitment to evolutionary fate in antibiotic resistance
  fitness landscapes.
\newblock Nature Communications. 2015;6:7385.

\bibitem{elena2007effects}
Elena SF, Wilke CO, Ofria C, Lenski RE.
\newblock Effects of population size and mutation rate on the evolution of
  mutational robustness.
\newblock Evolution. 2007;61:666--674.

\bibitem{otto2009evolutionary}
Otto SP.
\newblock The evolutionary enigma of sex.
\newblock The American Naturalist. 2009;174:S1--S14.

\bibitem{lynch1995mutational}
Lynch M, Conery J, Burger R.
\newblock Mutational meltdowns in sexual populations.
\newblock Evolution. 1995;49:1067--1080.

\bibitem{misevic2006sexual}
Misevic D, Ofria C, Lenski RE.
\newblock Sexual reproduction reshapes the genetic architecture of digital
  organisms.
\newblock Proceedings of the Royal Society of London B. 2006;273:457--464.

\bibitem{walker2012evolutionary}
Walker BL, Ofria C.
\newblock Evolutionary potential is maximized at intermediate diversity levels.
\newblock In: Adami C, Bryson DM, Ofria C, Pennock R, editors. Proceedings of
  the 13th International Conference on the Synthesis and Simulation of Living
  Systems. Cambridge, MA: MIT Press; 2012. p. 116--120.

\bibitem{zaman2014coevolution}
Zaman L, Meyer JR, Devangam S, Bryson DM, Lenski RE, Ofria C.
\newblock Coevolution drives the emergence of complex traits and promotes
  evolvability.
\newblock PLoS Biology. 2014;12:e1002023.

\bibitem{pennock2007models}
Pennock RT.
\newblock Models, simulations, instantiations, and evidence: the case of
  digital evolution.
\newblock Journal of Experimental \& Theoretical Artificial Intelligence.
  2007;19:29--42.

\bibitem{hindre2012new}
Hindr{\'e} T, Knibbe C, Beslon G, Schneider D.
\newblock New insights into bacterial adaptation through in vivo and in silico
  experimental evolution.
\newblock Nature Reviews Microbiology. 2012;10:352--365.

\bibitem{lenski1999genome}
Lenski RE, Ofria C, Collier TC, Adami C.
\newblock Genome complexity, robustness and genetic interactions in digital
  organisms.
\newblock Nature. 1999;400:661--664.

\bibitem{lenski2003evolutionary}
Lenski RE, Ofria C, Pennock RT, Adami C.
\newblock The evolutionary origin of complex features.
\newblock Nature. 2003;423:139--144.

\bibitem{goldsby2014evolutionary}
Goldsby HJ, Knoester DB, Ofria C, Kerr B.
\newblock The evolutionary origin of somatic cells under the dirty work
  hypothesis.
\newblock PLoS Biology. 2014;12:e1001858.

\bibitem{lenski1991long}
Lenski RE, Rose MR, Simpson SC, Tadler SC.
\newblock Long-term experimental evolution in Escherichia coli. I. Adaptation
  and divergence during 2,000 generations.
\newblock The American Naturalist. 1991;138:1315--1341.

\bibitem{ohta1996development}
Ohta T, Gillespie JH.
\newblock Development of neutral and nearly neutral theories.
\newblock Theoretical population biology. 1996;49:128--142.

\end{thebibliography}

\end{document}